\documentstyle [12pt]{article}

\def\aop{{\bf a}}
\def\nop{{\bf n}}
\def\Hop{{\bf H}}

\title{Dynamics of the Two-Site Hubbard Model}
\author{A. T. Costa Jr.\thanks{E-mail: antc@if.uff.br} 
and M. T. Thomaz\thanks{E-mail: mtt@if.uff.br}\\
{\it Instituto de F\'\i sica - Universidade Federal Fluminense}\\
{\it Av. Gal. Milton Tavares de Souza s/n.$\!\!^\circ$ - Gragoat\'a}\\
{\it Niter\'oi, Rio de Janeiro 24210-340, Brazil }}

\date{\today}

\begin{document}

\maketitle

\abstract{
The dynamical evolution of a two-site Hubbard model is derived in the
presence of an uniform external magnetic field for a general initial state. 
The time evolution of the half-filled (two-particle) case  has a
  complex behaviour. Under certain initial conditions, 
the average number of fermions in one-particle state, in the 
half-filled case, is not necessarily periodic, even though the
magnetization remains periodic. The results obtained may be applied
 to study the magnetization and the transition dipole moment of the 
organic charge-transfer salts in the in-phase mode.
}

\newpage

{\large{\bf Dynamics of the Two-Site Hubbard Model}}

\bigskip

Even the simplest fermionic many-body model in Condensed Matter
Physics, is too complicated to be
treated exactly. This is the case of the well known Hubbard model
\cite{hubbard}. Despite of its simplicity, it gives good qualitative
descriptions of many important phenomena in Condensed Matter Physics.
These results come from numerical simulations (of the Monte Carlo type,
for example) or from a perturbative treatment of one of the terms in the
hamiltonian \cite{hubbard}. Non-perturbative analytical treatment 
of the model turns out to be feasible if one further simplification
 is made: to consider a small number of space lattice wites. 
Although this simplification seems to impose
serious limitations on the applicability of the model, it still can
describe interesting phenomena occurring in real systems, e.g.,
 the appearance  of extra excitation lines in the infra-red spectra
 of organic charge-trasfer (CT) salts \cite{CTsalts, kozlov,torrence}. 
In the literature, this system is described by  two-site Hubbard 
model\cite{CTsalts, kozlov}. Usually, it has been studied 
energy-levels\cite{kozlov} and the optical properties of 
the CT salts in the presence of the external time dependent 
electric field\cite{CTsalts}, in the half-filled case. There are 
also experimental data of the energies of the peaks in the 
dimer spectrum for different temperature. In all previous work,
the system is not studied in the presence of an external magnetic 
field.

The time dependence of the CT salts results is a consequence of 
an external electric field. However, it is missing the dynamics of 
the CT salts under general initial conditions of the vector state
that describes the dimer.

\vspace{0.5cm}

In this work we present the exact time evolution of the two-site Hubbard
model, with arbitrary band filling and coupling constant, in the presence
of an uniform magnetic field. From an arbitrary many-body initial state 
we obtain the exact density operator of the system. The one-body properties
of this identical particles system are obtained by calculating 
the one-particle reduced density matrix. 
These results give us a dynamical picture of the
behaviour of the one-electron states of the system, and allow us 
to obtain the electron population in each one-particle state, the
magnetization per site and the transition of the electric dipole 
moment of CT salts in the in-phase mode, in the presence of an external
constant magnetic field.

The well-known Hubbard model \cite{hubbard} has the second-quantized 
hamiltonian

\begin{equation}
{\bf H}=\sum_{ij,\sigma}t_{ij}\aop^\dagger_{i,\sigma}\aop_{j,\sigma}+
U\sum_{i,\sigma}\nop_{i,\sigma}\nop_{i,-\sigma} + 
\lambda_B\sum_i (\nop_{i,\uparrow}-\nop_{i,\downarrow})\, .
\label{hamilt}
\end{equation}

\noindent  where $\aop_{j,\sigma}$ is the destruction operator of
one electron
with spin $\sigma$ on site $i$, $t_{ij}$ are the so-called hopping
integrals, $U$ is the effective intra-atomic Coulomb interaction, 
$\lambda_B=-{g\mu_B B \over 2}$, where $g$ is the Land\'e factor,
$\mu_B$ is the Bohr magneton, $B$ is the external magnetic 
field chosen in the $\hat{z}$ direction, and 
$\nop_{i,\sigma}=\aop^\dagger_{i,\sigma}\aop_{i,\sigma}$. 
The creation ($\aop_{i,\sigma}^\dagger$) and destruction 
($\aop_{j,\sigma^\prime}$)  operators satisfy the anti-commutation
relations: 

\begin{eqnarray}
\{ \aop^\dagger_{i,\sigma},\aop_{j,\sigma '}\} =
\delta_{ij}\delta_{\sigma\sigma '}\, , \nonumber \\
\label{commut}
\\
\{ \aop_{i,\sigma},\aop_{j,\sigma '}\} =
\{ \aop^\dagger_{i,\sigma},\aop^\dagger_{j,\sigma '}\} = 0\, . \nonumber
\end{eqnarray}

We consider here the case where the hamiltonian (\ref{hamilt}) takes into
account only nearest-neighbour hopping. We also 
have $t_{11} = t_{22} = E_0$ and $t_{12}= t_{21} = T$. Under these 
conditions, a two-site Hubbard hamiltonian (\ref{hamilt}) describes,
for example, the CT salts in the in-phase mode \cite{CTsalts}, 
but in absence of vibronic couping with the internal modes of
 the monomers. For the CT salt, $E_0$ denotes the enrgy of the
radical electron molecular orbital, $T$ gives the hopping integrals
of electrons  between  the molecules and $U$ is the effective
Coulomb interaction between the two elctrons on the same
 molecule\cite{kozlov}.

\vspace{0.5cm}

In order to study the most general state of the system for
 the two sites case,  we choose a basis of eigenstates of the number
 operator $\nop_{i,\sigma}$,  whose components can be written as

\begin{equation}
|n_1n_2;m_1,m_2\rangle = (\aop^\dagger_{1,\downarrow})^{n_1}
(\aop^\dagger_{1,\uparrow})^{n_2}(\aop^\dagger_{2,\downarrow})^{m_1}
(\aop^\dagger_{2,\uparrow})^{m_2}|0\rangle\, ,
\label{basis}
\end{equation}

\noindent where $n_i, m_i = 0,1$ and $i=1,2$. $|0\rangle$ represents the 
vacuum state of the model.

The action of $\Hop$ over each of the basis states can be obtained
by straightforward calculation, using the anticommutation 
relations eq. (\ref{commut}).

The  most general initial condition for the system described by a 
state vector is 

\begin{equation}
|\Psi (0)\rangle = \sum_{n_1,n_2,m_1,m_2=0}^1 f_{n_1,n_2,m_1,m_2}(0)
|n_1n_2;m_1m_2\rangle\, ,
\label{gen_st0}
\end{equation}

\noindent where   $|n_1n_2;m_1m_2\rangle$ are  given by eq.(\ref{basis})
and the values of the constants $f_{n_1,n_2,m_1,m_2}(0)$
are determined by the initial conditions. Due to the fact that the 
eigenstates of the number operator, eq. (\ref{basis}), form
a complete basis, at any time we have, 

\begin{equation}
|\Psi (t)\rangle = \sum_{n_1,n_2,m_1,m_2=0}^1 f_{n_1,n_2,m_1,m_2}(t)
|n_1n_2;m_1m_2\rangle\, .
\label{gen_st}
\end{equation}

Inserting $|\Psi (t)\rangle $ in the Schr\"odinger equation gives 
a system of coupled first order differential equations for the coefficients
$f_{n_1,n_2,m_1,m_2}(t)$ that can be solved analytically. Actually,
before solving the coupled equations for the coefficients, it is
 worth to notice that, since 
$\lbrack \Hop, {\bf N}_\sigma\rbrack=0$,
where ${\bf N}_\sigma =\sum_i\nop_{\sigma,i}$, 
 this system of coupled equations  brakes up in
 smaller set of systems. The dynamics of states with different  spin 
components are decoupled.
We end up with five  systems of first order differential equations.
In appendix A, we give the explicity time evolution  of each
eigenstate  of the total number operator
(${\bf N} = \sum_{i, \sigma}\nop_{\sigma,i}$). Not all 
vectors belonging to basis (\ref{basis}) are eigenstates 
of the  hamiltonian (\ref{hamilt}), but since {\bf N} and
${\bf N}_\sigma$ are constants of motion, the eigenstates of {\bf H}
are labeled  by their eigenvalues of  {\bf N} and 
${\bf N}_\sigma$.  
 
With the time-dependence of the most general state of the system in
 hands, we can write  the 
time-dependent density matrix of this many-body system,

\begin{equation}
\rho (t) = |\Psi (t)\rangle \langle\Psi (t) |\, .
\label{dens_matr}
\end{equation}

\noindent If we are interested in the effective dynamics of the
 one-particle subsystem,  the one-body density matrix \cite{feynman,ring} 
must be determined. It is defined in terms of the full density matrix as

\begin{equation}
\Lambda_{i, \sigma; j, \sigma^\prime} (t) =
 Tr\{ \aop^\dagger_{j, \sigma^\prime} \aop_{i, \sigma}\; \rho (t)\}\, .
\label{onebody_dm}
\end{equation}

For the sake of simplicity, instead of working with four indices in
the definition of $\Lambda (t)$ we redefine them as: 

\begin{equation}
{\bf a}_{1, \downarrow} \equiv {\bf a}_1,  \hspace{0.3cm}
{\bf a}_{2, \downarrow} \equiv {\bf a}_2,  \hspace{0.3cm}
{\bf a}_{1, \uparrow} \equiv {\bf a}_3,  \hspace{0.3cm}
{\bf a}_{2, \uparrow} \equiv {\bf a}_4,  
\end{equation}

\noindent and in an analogous way  the creation operators.

Using the new definition for the indices, and, the fact that $\rho (t)$
 is a  pure density matrix,  the trace in eq.(\ref{onebody_dm}) 
reduces to

\begin{equation}
\Lambda_{ij}(t) = \langle\Psi (t) | \aop^\dagger_j\aop_i |\Psi (t)\rangle\,
\hspace{1cm} i, j= 1, \cdots, 4 .
\label{simple_obdm}
\end{equation}

\noindent  It is clear from the definition (\ref{onebody_dm})
that the diagonal elements of $\Lambda(t)$ give the average population 
of the one-particle state $i$ for the identical particle 
system described by $|\Psi (t)\rangle$.
From these elements, other important quantities may be determined,
such as magnetization per site\footnote{ The $z$ component of the
 magnatization of site $i$ is\cite{kittel}

\begin {equation}
{\bf m}^{z}_i = - g \mu_B ({\bf n}_{i,\uparrow} - {\bf n}_{i,\downarrow}),
\end{equation}

\noindent where $g$ is the Land\'e factor and $\mu_B$ is the 
Bohr magneton.} and electric dipole moment\footnote{ The electron 
electric dipole moment of the dimer\cite{kittel}:

\begin {equation}
\vec{\bf p} = \frac{e \vec{a}}{2} ({\bf n}_1 - {\bf n}_2),
\end {equation}

\noindent where ${\bf n}_i = \sum_{ \sigma= \uparrow, \downarrow}
{\bf n}_{i,\sigma}$.  }. It is interesting
to notice that, although the whole system is in a pure state, the effective 
one-body subsystem, depending on the initial values of 
$f_{n_1,n_2,m_1,m_2}(0)$, is in a statistical mixture of states. This may be easily
verified by direct calculation of $\Lambda^2(t)$ and subsequent 
comparison with $\Lambda(t)$ itself. In general, we get 
$\Lambda^2(t)\neq \Lambda(t)$, in a clear
indication of the statiscal mixture character of the one-particle 
subsystem state. 

Although we have an analitical expression for the most general
state of the system $|\Psi(t)\rangle$, it is very cumbersome and not worth
showing here (look in Appendix A for the time dependence of
the coefficients $f_{m_1m_2n_1n_2}(t)$ eq. (\ref{gen_st})).
Instead, we choose a set of numerical values for the constants
of the model and some initial states $|\Psi(0)\rangle$, and present the 
results for some one-particle quantities.

The zero- and four-particle sectors of the Fock space are very 
uninteresting. They have only one vector each, which evolves in
 time only by a phase. If the initial state $|\Psi(0)\rangle$ lies 
entirely in one of these sectors, all the one-particle observables
 are constant in time.

The one- and three-particle sectors have both four vectors, and are divided 
into two sets of two coupled vectors each. If $|\Psi(0)\rangle$ lies within
one of these sectors, the one-particle observables are time-dependent 
in general, being periodic functions of time with the frequency
$\frac{T}{\hbar}$.

Since the dynamics does not couple different Fock sub-spaces, the diagonal
elements of the $\Lambda$ matrix in these two Fock-subspaces are:

\begin{eqnarray}
\Lambda^{(1)}_{11} & = &  
\mid f(1,0;0,0;0)\mid^2 \cos^2 (\frac{Tt}{\hbar}) +
\mid f(0,0;1,0;0)\mid^2 \sin^2 (\frac{Tt}{\hbar}), \nonumber \\
\label{N1.1} \\
\nonumber\\
\Lambda^{(1)}_{33} & = &  \mid f(1,0;1,1;0)\mid^2 + \mid f(1,1;1,0;0)\mid^2+
\nonumber \\
 & + & \mid f(1,1;0,1;0)\mid^2 \cos^2 (\frac{Tt}{\hbar}) +
\mid f(0,1;1,1;0)\mid^2 \sin^2 (\frac{Tt}{\hbar}) \nonumber \\
  \label{N1.2}
\end{eqnarray}

\noindent and 

\begin{eqnarray}
\Lambda^{(3)}_{11} & = &  
\mid f(0,1;0,0;0)\mid^2 \cos^2 (\frac{Tt}{\hbar}) +
\mid f(0,0;0,1;0)\mid^2 \sin^2 (\frac{Tt}{\hbar}), \nonumber \\
\label{N3.1} \\
\Lambda^{(3)}_{33} & = & \mid f(0,1;1,1;0)\mid^2 + \mid f(1,1;0,1;0)\mid^2+
\nonumber \\
 & + & \mid f(1,1;1,0;0)\mid^2 \cos^2 (\frac{Tt}{\hbar}) +
\mid f(1,0;1,1;0)\mid^2 \sin^2 (\frac{Tt}{\hbar}), \nonumber \\ \,
  \label{N3.2}
\end{eqnarray}

\noindent where $\Lambda^{(1)}_{ii}$ are diagonal elements of the
one-particle reduced density matrix in the Fock sub-space $N=1$ and
 $\Lambda^{(3)}_{ii}$ the diagonal elements for $N=3$.
In the solid-state language,
these two sectors correspond to the quarter and three-quarter band filling
of the model respectively. Due to the particle-hole symmetry of the model, these two sectors have very similar characteristics. Note that when the state 
has three fermionic particles, there is interaction among them. However,
we see from eqs. (\ref{N3.1}) and (\ref{N3.2}) that the dynamics
 of theses states is only dictated by the the hopping integral $T$,
as is compatible  with particle-hole symmetry. 

The operator ${\bf N}_\sigma$ is a constant of motion. To describe the
dynamics of the one-particle sub-system it is enough to discuss,
 for fixed $\sigma$, the population in one of the space sites.
 We have chosen to fix in eqs. (\ref{N1.1})-({\ref{N1.2}) and 
eqs.(\ref{N3.1})-(\ref{N3.2}) on site $i=1$. 

The two-particle sector has the richest structure, for it has six
 state vectors, where  four of them are coupled. If we choose 
$|\Psi(0)\rangle$ to lie in this sector, the 
one-particle observables are in general quasi-periodic functions of time.
The time dependence of the population of each one-particle observable has
two frequencies: $ w_1 = \frac{U}{2\hbar}$ and \break
$w_2 = \frac{\sqrt{ U^2 + 16 T^2}}{2\hbar}$. In general, 
$\frac{w_1}{w_2} \not= \frac{p}{q}$, where $p$ and $q$ are
integers, which implies that the dynamics of the population 
 of the one-particle sub-system is not periodic in general. 
For some special initial states, only a few of the one-particle observables 
turn out to be periodic. This happens because of subtle
cancellations occurring in the calculations of those observables, as will 
be shown later on. This sector correspond to the half-filled Hubbard model,
widely studied in solid state physics. In particular, this is the
band-filling used to describe the charge-trasfer
 organic salts \cite{CTsalts, kozlov}.

In order to ilustrate the qualitative discussion given above, we
present some graphs for one-particle observables obtained from
the analytical expressions. We need to point out that the
 external constant magnetic field only appears in the 
phases of the time evolution of states (\ref{basis}) of the 
Fock sub-spaces $N=1$, $N=3$ and for states of $N=2$ when both 
fermions have the same spin component. But, in the one-particle
reduced density matrix, it gives no contribution. 
Let us consider the graphs when the model  constants are chosen as:

\begin{eqnarray}
  T &=& 1 \nonumber\\
U &=& 0.5\, , \nonumber \\
\lambda_B & =& 0\, .
\label{const1}
\end{eqnarray}

\noindent  We have taken $\hbar=1$. In doing this, $T/\hbar$ 
turns out to be an energy scale for the system. We first chose
$U$ comparable to the hopping integral $T$ to see how the
two terms in the Hamiltonian combine when none of them is 
dominating. This competition  reveals itself in the time-dependence 
of the one-particle observables by the absence of a 
characteristic frequency in their oscillations. In figure \ref{fig1}
we show the average spin up occupation $n_{1\uparrow}=
\Lambda_{1\uparrow,1\uparrow}\equiv \Lambda_{33}(t)$, 
and the average electric dipole moment of the dimmer
$d(t)=\Lambda_{2\uparrow,2\uparrow}(t)+
\Lambda_{2\downarrow,2\downarrow}(t)-(\Lambda_{1\uparrow,1\uparrow}(t)+
\Lambda_{1\downarrow,1\downarrow}(t)) \equiv \Lambda_{22}(t) + 
\Lambda_{44}(t)-(\Lambda_{33}(t) + 
\Lambda_{11}(t))$
for the initial state $|\Psi_1(0)\rangle=|11\rangle|00\rangle$. 
It is easy to identify both functions as non-periodic by simply looking
at the figures. The average magnetization per site of this state is zero, 
since  its initial value is zero and it is a constant of motion.

If we now choose a new initial state $|\Psi_2(0)\rangle=\frac{1}{\sqrt{3}}
(|10\rangle|01\rangle -|01\rangle|10\rangle +|11\rangle|00\rangle)$, we
obtain two non-periodic observables, namely 
$n_{1\uparrow}(t)$ and $d_1(t)$, but
the average magnetization of site $1$, 
$m_1(t)=\Lambda_{1\uparrow,1\uparrow}(t)-
\Lambda_{1\downarrow,1\downarrow}(t)$, turns out to be
periodic ( see Figure \ref{fig2}). 
This occurs because of a cancellation of the contributions from 
$|10\rangle|01\rangle$ and $|01\rangle|10\rangle$ to the
average magnetization per site, due to the fact that both states
enter the combination with coefficients of same modulus but opposite sign.

In order to ilustrate how the relation between the hopping integral $T$ 
and the Coulomb interaction $U$ may influence the observables of the system,
we choose a new set of model constants, and repeat the calculations already
performed. In the literature\cite{torrence} it is stated that 
for the CT salts $U= 5.45 T$, therefore, we choose the
 new set of constants as:

\begin{eqnarray}
 T &=& 1 \nonumber\\
U &=& 5.45\, , \nonumber \\
\lambda_B &=& 0\, .
\label{const2}
\end{eqnarray}

\noindent Now the Coulomb interaction $U$ is four times greater
 than the hopping integral $T$. In figure \ref{fig3} we observe 
that this noticeable  difference between them causes the appearance 
of  a modulation of frequencies. The higher frequency oscilations
due to the Coulomb integral are superimposed to the hopping integral 
oscilations.

\vspace{0.5cm}

In summary, we studied the exact dynamics of two-site Hubbard model
 when the identical particles system is 
described by a vector state. We considered the case where 
$t_{12}= t_{21} = T$, that has been used as a model to explain 
the CT salts in the in-phase mode \cite{CTsalts}, without the interaction
of the radical electrons with the internal vibronic modes
of the molecule.

The particle--hole simmetry of the Hubbard model is well 
realized by the time evolution of $\Lambda^{(3)}_{ii}$, $i=1,2$. 
Even though for the three-particle Fock sub-space the
Coulomb interaction (\ref{hamilt}) still works, the
dynamics of the population of one-particle states is governed
by the hopping integral $T$, as happens for the Fock sub-space 
$N=1$. 

In the half-filled case, depending on the initial conditions, 
we can get non-periodic behaviour for some one-particle properties, while 
the magnetization per site remains periodic.

Finally, we have shown as the one-particle observables as magnetization
 and site occupation oscillate in time. It is well known that, when these
quantities are measured in real systems as magnetic transition metals for
example, the measurements furnish constant fractional values. 
From the  theory of intinerant magnetism we have that
the measured fractional charge and magnetization of such metals is a
consequence of a time average performed by the measuring apparatus.
 We  obtain  fractional values by taking the
time average of the time-dependentt observables we have calculated. This
time average is defined as

\begin{equation}
\langle w(t)\rangle = \lim_{T\rightarrow\infty}\frac{1}{T}\int_0^T w(t)dt\, .
\label{time_av}
\end{equation}

For the half-filled case, we get  from  eq.(\ref{time_av}) and what 
was discussed previously, that the average value for the magnetization
per site is a fractional constant, but the eletric dipole transition is
 not, since it is not a periodic function in time. 
We get a time dependent electric polarizability in the absence  of 
an external electric field. It is an open question if this effect can actually be measured.

\vspace{1cm}

{\large{\bf Acknowledgements} }

\bigskip 

A.T.C. Jr.  thanks CNPq for financial support and
M.T.T. thanks CNPQ and FINEP for partial financial support. We also 
thank J.J. Rodriguez Nu\~nes for bringing reference \cite{kozlov} to
our attention.

\newpage

\noindent  {\large\bf Appendix A: Time Evolution of the Eigenstates of 
${\bf n}_{i,\sigma}$  Operator }
\bigskip

In this apendix we present the explicit time dependence of the most 
general state  of the two-site Hubbard model. It is obtained by inserting 

\begin{equation}
|\Psi(t)\rangle = \sum_{m_1,m_2,n_1,n_2=0}^1 
f_{m_1m_2n_1n_2}(t)|m_1m_2\rangle |n_1n_2\rangle
\label{most_gen}
\end{equation}  

\noindent  in the Schr\"odinger equation, and then projecting out 
this equation over each one of the basis states (\ref{basis}). 
This will, in 
principle, give a set of sixteen coupled  first order 
differential equations for the $f_{m_1m_2n_1n_2}(t)$. Due to some
simetries of the Hamiltonian, these sixteen equations decouple,
 forming a set of five independent systems with variable number
 of equations. From the hamiltonian (\ref{hamilt}), we have
the folling constants of motion:

\begin{eqnarray}
\lbrack \Hop ,{\bf N}_{\sigma}\rbrack &=& 0\, ,\nonumber\\
\lbrack \Hop ,{\bf N}\rbrack &=& 0\, . 
\label{simetr}
\end{eqnarray}

\noindent  Eq.(\ref{simetr}) leads to total charge and
 spin conservation of the hamiltonian, which defines five
 independent sectors of the Hilbert space, labeled by the total 
charge of the states. These sectors may be further divided in subsectors
of states with the same total spin (or magnetization). Therefore 
we may determine the dynamics of each subsector independently. The 
explicit time dependence of each coefficient $f_{m_1m_2n_1n_2}(t)$ 
is given below.

\begin{eqnarray}
& &\mbox{\bf Fock sub-space N=0:}\nonumber \\
& &f_{0000}(t) = f_{0000}(0)\\
 \nonumber\\
& &\mbox{\bf Fock sub-space N=1:}\nonumber \\
& &\mbox{\rm  Subsector $\sigma= \downarrow$:}\nonumber \\
& &f_{1000}(t) = \lbrack f_{1000}(0)cos(\frac{t_{12}t_{21}}{\hbar}t)-
i\frac{|\tau | }{\tau^*}
f_{0010}(0)sin(\frac{t_{12}t_{21}}{\hbar}t)\rbrack e^{-i\frac{(\epsilon - 
\lambda_B)}{\hbar}t}\\
& &f_{0010}(t) = \lbrack f_{0010}(0)cos(\frac{t_{12}t_{21}}{\hbar}t)-
i\frac{|\tau | }{\tau}
f_{1000}(0)sin(\frac{t_{12}t_{21}}{\hbar}t)\rbrack e^{-i\frac{(\epsilon -
 \lambda_B)}{\hbar}t}\\
\nonumber\\
& &\mbox{Subsector $\sigma= \uparrow$:}\nonumber\\ 
& &f_{0100}(t) = \lbrack f_{0100}(0)cos(\frac{t_{12}t_{21}}{\hbar}t)-
i\frac{|\tau | }{\tau^*}
f_{0001}(0)sin(\frac{t_{12}t_{21}}{\hbar}t)\rbrack e^{-i\frac{(\epsilon - 
\lambda_B)}{\hbar}t}\\
& &f_{0001}(t) = \lbrack f_{0001}(0)cos(\frac{t_{12}t_{21}}{\hbar}t)-
i\frac{|\tau | }{\tau}
f_{0100}(0)sin(\frac{t_{12}t_{21}}{\hbar}t)\rbrack e^{-i\frac{(\epsilon - 
\lambda_B)}{\hbar}t}\\
\nonumber\\
& &\mbox{\bf Fock sub-space N=2:}\nonumber\\
& &\mbox{Subsector $\sigma=0$:}\nonumber\\
& &f_{1100}(t) = i\frac{|\tau |^2}{\tau^*\beta}sin(\frac{\beta}{\hbar}t)
(f_{0110}(0)-f_{1001}(0))
+ \frac{1}{2\beta\tau^*}\lbrack \beta cos(\frac{\beta}{\hbar}t)
 -iUsin(\frac{\beta}{\hbar}t)\rbrack\times\nonumber\\
& &\times(\tau f_{0011}(0)+\tau^* f_{1100}(0) )
e^{-i\frac{(2\epsilon+U)}{\hbar}t}
-\frac{1}{2\tau^*}e^{-2i\frac{(\epsilon+U)}{\hbar}t}
(\tau f_{0011}(0)-\tau^* f_{1100}(0) ) \nonumber\\
		\\
\nonumber\\
& &f_{0011}(t) = i\frac{|\tau |^2}{\tau\beta}sin(\frac{\beta}{\hbar}t)
(f_{0110}(0)-f_{1001}(0))
+ \frac{1}{2\beta\tau}\lbrack \beta cos(\frac{\beta}{\hbar}t)
 -iUsin(\frac{\beta}{\hbar}t)\rbrack\times\nonumber\\
& &\times(\tau f_{0011}(0)+\tau^* f_{1100}(0) )
e^{-i\frac{(2\epsilon+U)}{\hbar}t}
+\frac{1}{2\tau}e^{-2i\frac{(\epsilon+U)}{\hbar}t}(\tau f_{0011}(0)
-\tau^* f_{1100}(0) )  \nonumber\\
		\\
\nonumber\\
& &f_{1001}(t) = \frac{1}{2}(f_{0110}(0)+f_{1001}(0))
e^{-2i\frac{\epsilon}{\hbar}t}
-\{ \frac{i}{\beta}\lbrack \tau f_{0011}(0)+\tau^* f_{1100}(0) +
 			\nonumber\\
& &+\frac{U}{2}(f_{0110}(0)-f_{1001}(0)\rbrack
sin(\frac{\beta}{\hbar}t)-\nonumber\\
& &-\frac{1}{2}(f_{0110}(0)-f_{1001}(0)) cos(\frac{\beta}{\hbar}t)\}
e^{-i\frac{(2\epsilon +U)}{\hbar}t}\\
\nonumber\\
& &f_{0110}(t) = \frac{1}{2}(f_{0110}(0)+f_{1001}(0))
e^{-2i\frac{\epsilon}{\hbar}t}
+ \{ \frac{i}{\beta}\lbrack \tau f_{0011}(0)+\tau^* f_{1100}(0) + 
			\nonumber\\
& &+\frac{U}{2}(f_{0110}(0)-f_{1001}(0)\rbrack
sin(\frac{\beta}{\hbar}t)-\nonumber\\
& &-\frac{1}{2}(f_{0110}(0)-f_{1001}(0)) cos(\frac{\beta}{\hbar}t)\} 
e^{-i\frac{(2\epsilon +U)}{\hbar}t} \\
\nonumber\\
& &\mbox{Subsector $\sigma= -1$:}\nonumber\\
& &f_{1010}(t) = e^{-2i\frac{(\epsilon-\lambda_B)}{\hbar}t}f_{1010}(0)\\
\nonumber \\
& &\mbox{Subsector $\sigma= 1$:}\nonumber\\
& &f_{0101}(t) = e^{-2i\frac{(\epsilon+\lambda_B)}{\hbar}t}f_{0101}(0)\\
\nonumber \\
& &\mbox{\bf Fock sub-space N=3:}\nonumber\\
& &\mbox{Subsector $\sigma= \downarrow$:}\nonumber\\
& &f_{1110}(t) = \lbrack i\frac{|\tau |}{\tau^*}
f_{1011}(0)sin(\frac{|\tau |}{\hbar}t)
+f_{1110}(0)cos(\frac{|\tau |}{\hbar}t)\rbrack 
e^{-i\frac{(3\epsilon +2U -\lambda_B)}{\hbar}t}\\
\nonumber\\
& &f_{1011}(t) = \lbrack i\frac{|\tau |}{\tau}f_{1110}(0)
sin(\frac{|\tau |}{\hbar}t)
+f_{1011}(0)cos(\frac{|\tau |}{\hbar}t)\rbrack 
e^{-i\frac{(3\epsilon +2U -\lambda_B)}{\hbar}t}\\
\nonumber \\
& &\mbox{Subsector $\sigma= \uparrow$:}\nonumber\\
& &f_{1101}(t) = \lbrack i\frac{|\tau |}{\tau^*}f_{0111}(0)
sin(\frac{|\tau |}{\hbar}t)
+f_{1101}(0)cos(\frac{|\tau |}{\hbar}t)\rbrack 
e^{-i\frac{(3\epsilon +2U +\lambda_B)}{\hbar}t}\\
\nonumber\\
& &f_{0111}(t) =\lbrack i\frac{|\tau |}{\tau}f_{1101}(0)
sin(\frac{|\tau |}{\hbar}t)
+f_{0111}(0)cos(\frac{|\tau |}{\hbar}t)\rbrack 
e^{-i\frac{(3\epsilon +2U +\lambda_B)}{\hbar}t}\\
\nonumber\\
& &\mbox{\bf Fock sub-space N=4:}\nonumber\\
& &f_{1111}(t) = e^{-4i\frac{(\epsilon+U)}{\hbar}t}f_{1111}(0)
\end{eqnarray}

\noindent  Here,  $f_{m_1m_2n_1n_2}(0)$ is the coefficient of 
the $|m_±m_2\rangle|n_1n_2\rangle$ state at $t=0$, 
$\tau = t_{12} = t_{21}^*$ and $\beta=\sqrt{U^2+4|\tau |^2}$.

\newpage

\begin{figure}
\caption{Average occupation of site 1 (a) and electric dipole moment (b)
for the first set of chosen constants (\protect{\ref{const1}})
and initial state $|\Psi_1(0)\rangle = |11\rangle |00\rangle$.}
\label{fig1}
\end{figure}

\begin{figure} 
\caption{Average occupation (a) and magnetization (b) of site 1 and electric dipole
moment (c) for the first set of chosen constants and 
initial state $|\Psi_2 (0)\rangle $. Observe the non periodicity
of (a) and (c), in contrast to (b), which is clearly periodic.}
\label{fig2}
\end{figure}

\begin{figure} 
\caption{Average occupation of site 1(a) and electric dipole
moment(b) for the second set of chosen constants 
(\protect{\ref{const2}}) 
and initial state $|\Psi_2(0)\rangle$. Contrary to what was seen
in figs. \protect{\ref{fig1}} and \protect{\ref{fig2}}, these 
observables present two clearly distinct characteristic frequencies 
superimposed, due to the discrepancy between the hopping integral and the 
Coulomb interaction.}
\label{fig3}
\end{figure}

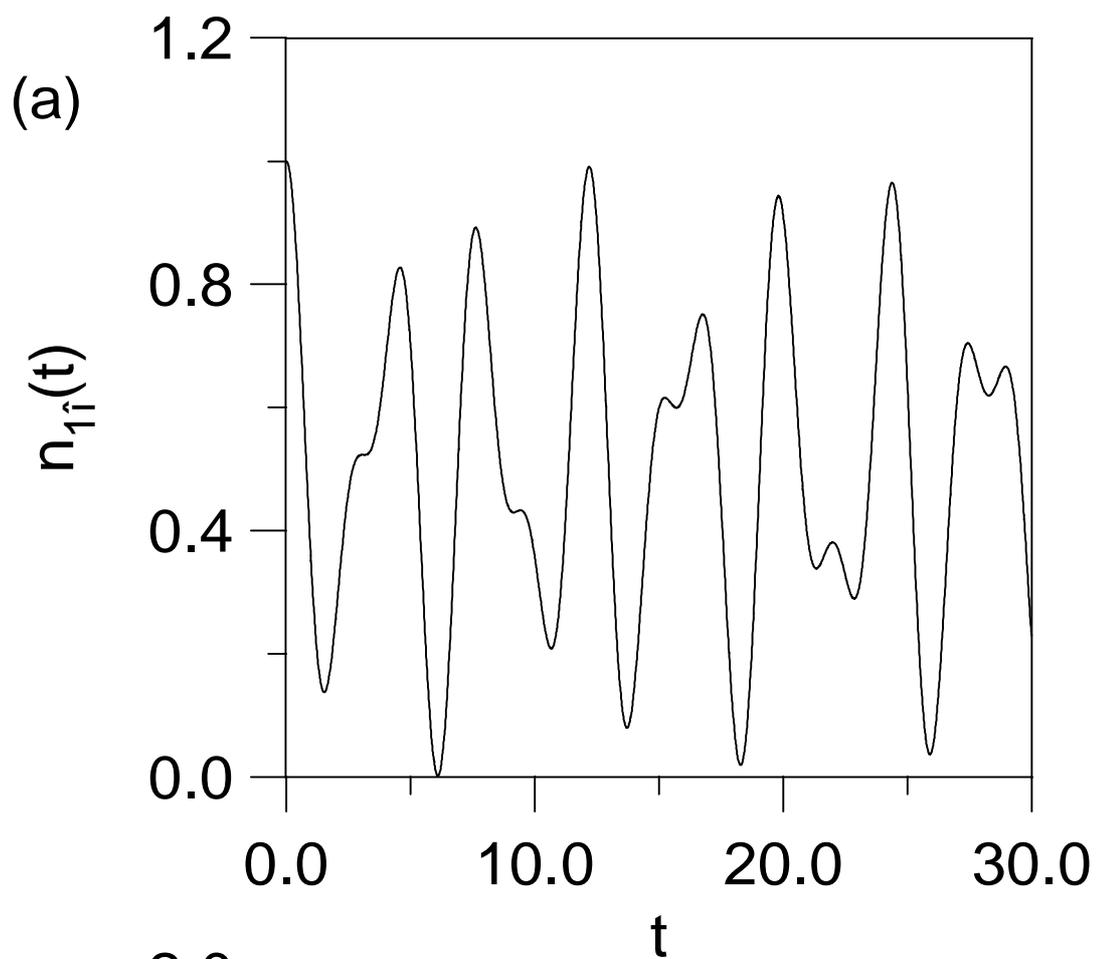
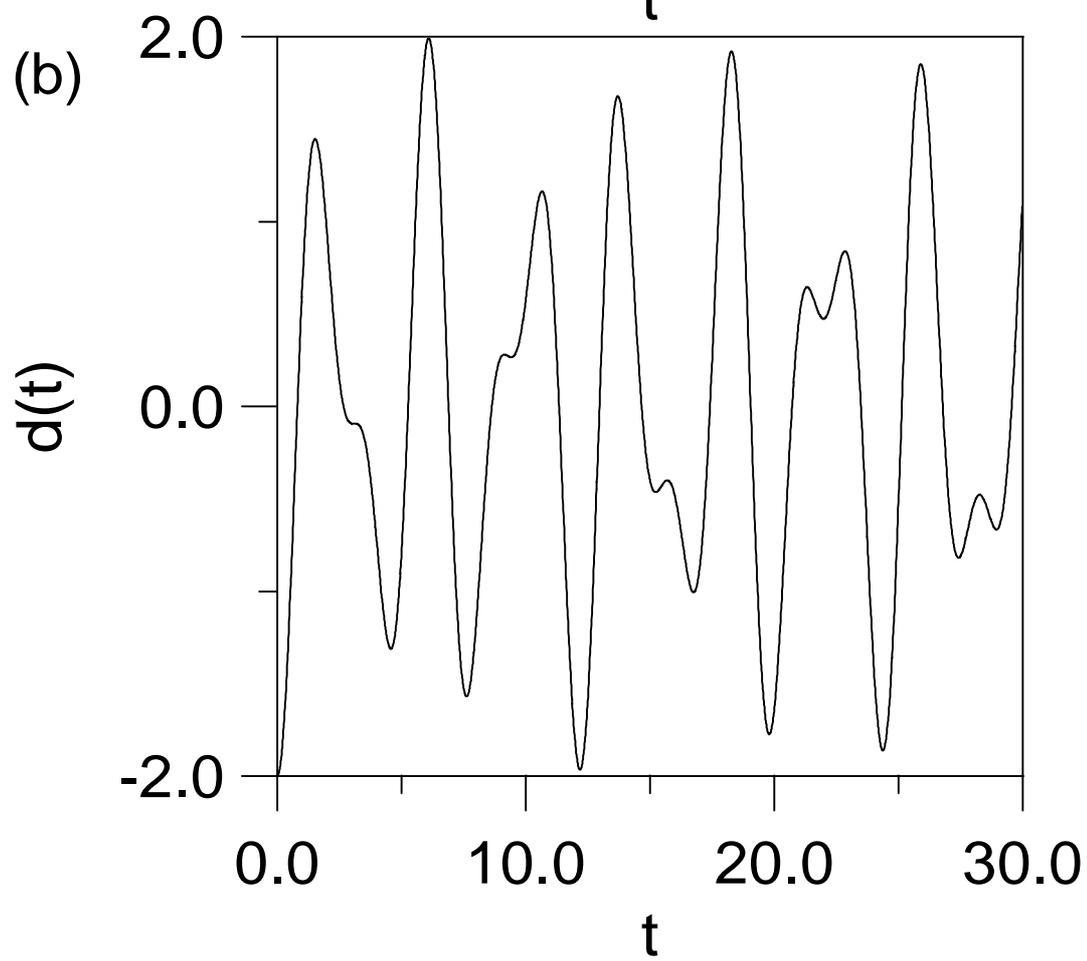
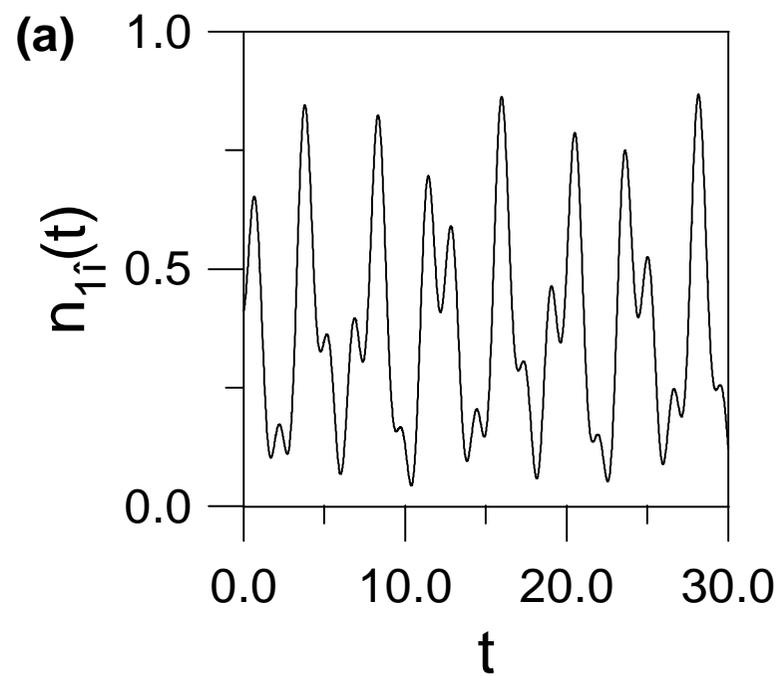
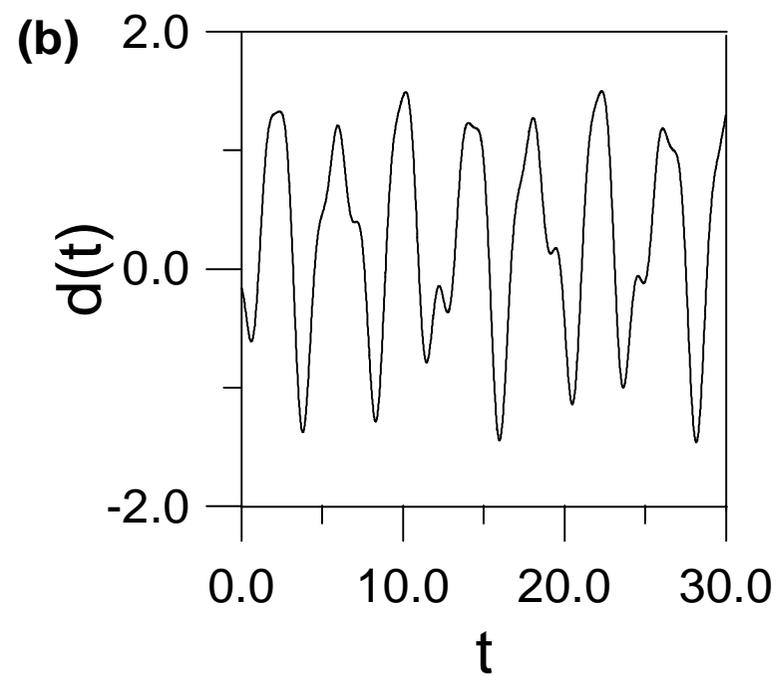
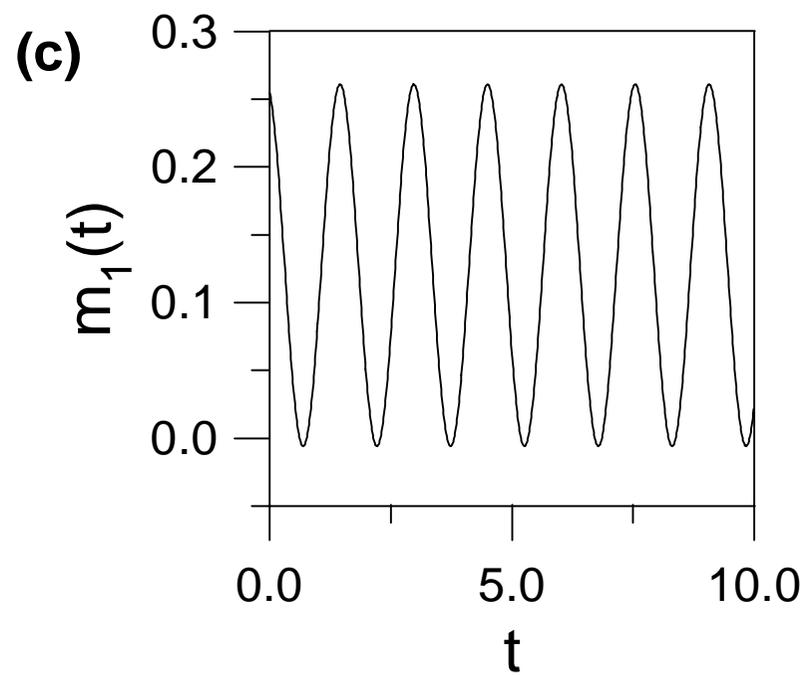
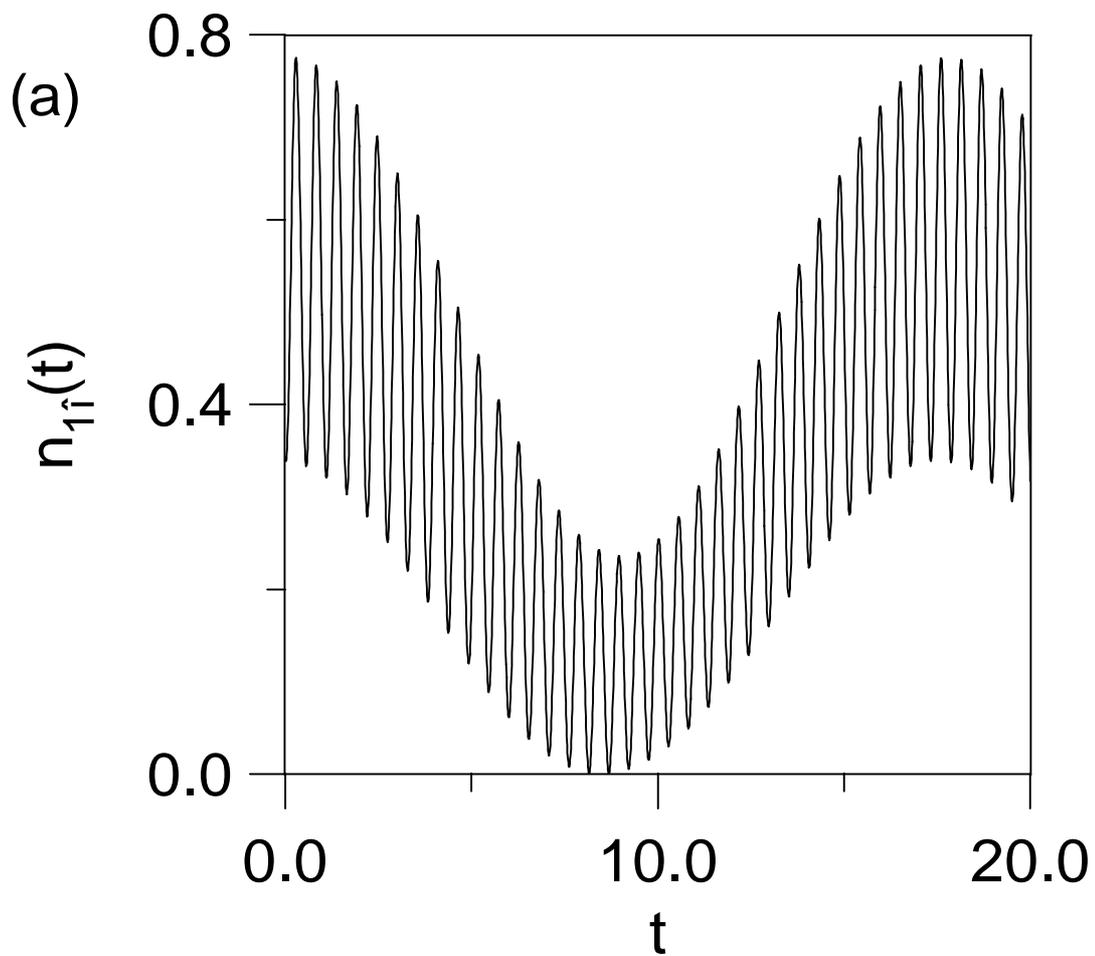
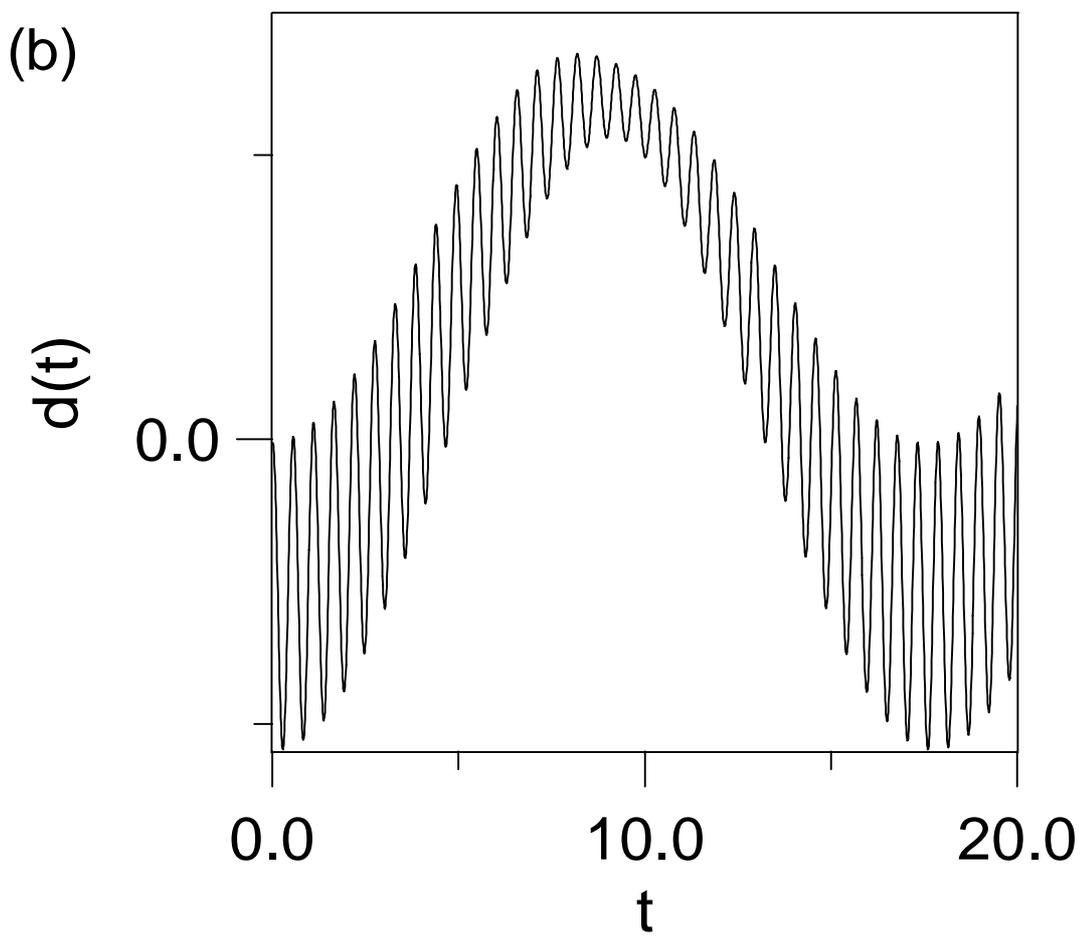
\end{document}